\documentclass[twocolumn,showpacs,preprintnumbers,amsmath,amssymb]{revtex4}
\usepackage{graphicx}
\usepackage{color}
\usepackage{epsfig}
\usepackage{latexsym}
\usepackage{bm}

\begin{document}

\renewcommand{\ni}{{\noindent}}
\newcommand{\dprime}{{\prime\prime}}
\newcommand{\be}{\begin{equation}}
\newcommand{\ee}{\end{equation}}
\newcommand{\bea}{\begin{eqnarray}} 
\newcommand{\eea}{\end{eqnarray}}
\newcommand{\nn}{\nonumber} 
\newcommand{\bk}{{\bf k}}
\newcommand{\bQ}{{\bf Q}}
\newcommand{\q}{{\bf q}}
\newcommand{\s}{{\bf s}}
\newcommand{\bN}{{\bf \rnabla}}
\newcommand{\bA}{{\bf A}}
\newcommand{\bE}{{\bf E}}
\newcommand{\bj}{{\bf j}}
\newcommand{\bJ}{{\bf J}}
\newcommand{\bs}{{\bf v}_s}
\newcommand{\bn}{{\bf v}_n}
\newcommand{\bv}{{\bf v}} 
\newcommand{\la}{\langle}
\newcommand{\ra}{\rangle} 
\newcommand{\dg}{\dagger}
\newcommand{\br}{{\bf{r}}} 
\newcommand{\brp}{{\bf{r}^\prime}} 
\newcommand{\bq}{{\bf{q}}}
\newcommand{\hx}{\hat{\bf x}} 
\newcommand{\hy}{\hat{\bf y}}
\newcommand{\bS}{{\bf S}} 
\newcommand{\cU}{{\cal U}}
\newcommand{\cD}{{\cal D}} 
\newcommand{\bR}{{\bf R}}
\newcommand{\pll}{\parallel} 
\newcommand{\sumr}{\sum_{\vr}} 
\newcommand{\cP}{{\cal P}} 
\newcommand{\cQ}{{\cal Q}} 
\newcommand{\cS}{{\cal S}}
\newcommand{\ua}{\uparrow} 
\newcommand{\da}{\downarrow}
\newcommand{\red}{\textcolor {red}}

\title{Non-equilibrium fluctuation theorems in the presence of local heating}

\author{Punyabrata Pradhan, Yariv Kafri, and Dov Levine}

\affiliation{ Physics Department, Technion, Haifa, Israel}

\begin{abstract}
\noindent
We study two non-equilibrium work fluctuation theorems, the Crooks' theorem and the Jarzynski equality, for a test system coupled to a spatially extended heat reservoir whose degrees of freedom are explicitly modeled. The sufficient conditions for the validity of the theorems are discussed in detail and compared to the case of classical Hamiltonian dynamics. When the conditions are met the fluctuation theorems are shown to hold despite the fact that the immediate vicinity of the test system goes out of equilibrium during an irreversible process. We also study the effect of the coupling to the heat reservoir on the convergence of $\langle \exp(-\beta W) \rangle$ to its theoretical mean value, where $W$ is the work done on the test system and $\beta$ is the inverse temperature.  It is shown that the larger the local heating, the slower the convergence.
\end{abstract}


\maketitle

\section{Introduction}

The recently discovered non-equilibrium fluctuation theorems deal with large deviations and symmetries of quantities \cite{review, HarrisShutz07} such as entropy production \cite{Evans_Cohen_Gallavotti, Lebowitz}, heat exchange \cite{JarzynskiPRL2004, SaitoPRL2007}, or work performed \cite{JarzynskiPRL1997, CrooksPRE1999} during an irreversible process. Two examples which have received much attention are the Jarzynski equality and the Crooks' theorem. In the Jarzynski equality one considers a test system (TS) which is initially in  equilibrium with a heat reservoir at temperature $T$. The test system is then driven to another state by varying an external parameter $\lambda$ over a finite time interval $0 \leq t \leq \tau$ using a fixed protocol $\lambda(t)$.  In this process, a history dependent work, $W$, is performed. The Jarzynski equality \cite{JarzynskiPRL1997} states that the average of $e^{-\beta W}$ over histories satisfies
\begin{equation}
\langle e^{-\beta W} \rangle \; \; = \; \; e^{-\beta \Delta F_{TS}}  \;\;,
\label{eq:JE}
\end{equation}
where $\beta=1/T$ is the temperature of the system when it is in equilibrium and $\Delta F_{TS} = F_{TS}({\cal B}) - F_{TS}({\cal A})$ is the free energy difference between the equilibrium states, $\cal A$ and $\cal B$, of the test system at the initial and final values of $\lambda$, respectively. 

The Crooks' theorem is a statement about the probability distribution of work performed in an irreversible process \cite{CrooksPRE1999}. As for the Jarzynski equality, one considers a test system in contact with a heat reservoir at temperature $T$.  Initially in an equilibrium state $\cal A$ with $\lambda = \lambda(0)$, the test system is driven, during the time interval $(0,\tau)$, by the protocol $\lambda(t)$ (the `forward process'). During this process, varying amounts of work may be performed, depending on the trajectory the system takes in state-space as $\lambda$ is varied; the likelihood that work $W$ is performed is specified by a probability distribution $P_F(W)$. One then considers the reverse process in which the system starts in the equilibrium state $\cal B$ with $\lambda = \lambda(\tau)$, and is driven by the time-reversed protocol $\tilde{\lambda}(t)\equiv \lambda(\tau-t)$ in the time interval $(0, \tau)$. In this process the probability of work $W$ being done is specified by the distribution $P_R(W)$. The Crooks' theorem states that 
\begin{equation}
\frac{P_F(W)}{P_R(-W)} = e^{\beta(W-\Delta F_{TS})} \;\;\;\; .
\label{eq:CT}
\end{equation}
Note that the Crooks' theorem is a stronger statement than the Jarzynski equality; indeed the latter can be obtained by integrating the former \cite{CrooksPRE1999}.  

As has been discussed extensively in the past \cite{JarzynskiPRL1997, JarzynskiPRE1997, grosberg}, both relations rely on rare events in which the work performed deviates significantly from the average.
Thus, the relevance of Eqs. (\ref{eq:JE}) or (\ref{eq:CT}) as a tool for measuring free-energy differences is restricted to small systems, for which the probability of sampling atypical events is non-negligible even for experimentally reasonable realizations of the protocol \cite{Liphardt}.

Since their discovery a decade ago, the fluctuation theorems have been proven for both stochastic \cite{JarzynskiPRE1997, CrooksJStatPhys1998, CrooksPRE1999} and Hamiltonian dynamics \cite{JarzynskiJStatMech2004, Cleuren_Broeck_Kawai, Park}. Although numerous proofs have been given, the precise assumptions necessary for their validity has been the subject of debate \cite{CohenJStatMech, CohenMolPhys, Palmieri}. In large measure, this lack of clarity revolves around the importance of {\it local heating}, to wit, that near the region of contact with the test system the heat reservoir is {\it itself} driven out of equilibrium during the experiment.  
For example, proofs relying on Markovian dynamics \cite{CrooksPRE1999} assume that the test system obeys detailed balance with an ideal heat reservoir, that is, the transition rate, $w^{TS}_{j,i}$  from configurations $i$ to $j$ of the test system satisfies
\begin{equation} 
w^{TS}_{j,i} \exp(-\beta E_i) = w^{TS}_{i,j} \exp(-\beta E_j) \;,
\label{CanonicalDB}
\end{equation}
where $E_i$ and $E_j$ are energies of the configurations $i$ and $j$ of the test system, respectively. Similarly, for Langevin dynamics one typically assumes white noise satisfying the fluctuation-dissipation relation \cite{Markov-dynamics}. Thus, these dynamics implicitly assume that the heat reservoir is always in thermal equilibrium \cite{Klein}. While local heating may, in practice, be small, it is nonetheless a fundamental issue which goes to the heart of the validity of the fluctuation theorems \cite{Blythe}.

The aim of this paper is to elucidate the requirements for the validity of the Crooks' theorem and the Jarzynski equality in the presence of local heating.  While several work have considered a restricted class of non-Markovian sources of noise acting as the heat reservoir \cite{Non-Markovian}, we take a different approach. To this end, we consider the test system, on which the driving is performed, to be a portion of a larger isolated ``combined system'' consisting of the test system + heat reservoir, with no assumption made about the heat reservoir degrees of freedom.  We first review results for classical Hamiltonian dynamics and their limitations, and then present a new proof for Markovian stochastic dynamics which requires less stringent restrictions than their classical Hamiltonian counterparts - specifically for classical Hamiltonian dynamics the motion in phase space is incompressible while for stochastic dynamics there is no such restriction.  We show that  in order for Crooks' theorem to hold, it is sufficient that the following three conditions are satisfied:
\begin{enumerate}
\item The closed ``combined system'' (CS), consisting of the test system + heat reservoir, obeys time-reversal symmetric dynamics when the external parameters (such as $\lambda$) are constant.
\item When in thermal equilibrium, extensive properties of the systems are additive, {\it e.g.}, the free energy of the CS is the sum of the free energies of the test system and the heat reservoir: $F_{CS} = F_{TS}+F_{B}$.
\item The control parameters (such as $\lambda$) directly couple only to the test system (though their influence may surely be felt in the entire CS).
\end{enumerate}
We note that when condition (2) does not hold, modified relations may still be obtained; we comment on this in the body of the paper.
Since we make no assumptions about the state of the heat reservoir, this demonstrates that local heating invalidates neither the Jarzynski equality nor the Crooks' theorem.  This notwithstanding, local heating {\it strongly impacts} on the statistics needed to accurately estimate free energy differences, and we address this issue numerically. 

\section{Hamiltonian Dynamics}

In the following we present a proof of the Crooks' theorem for an isolated system obeying classical Hamiltonian dynamics, along the lines of Reference \cite{Cleuren_Broeck_Kawai}. This will allow us to emphasize the differences between stochastic and classical Hamiltonian dynamics.  We wish to stress that for closed systems (such as the CS above) obeying classical Hamiltonian dynamics, the Crooks' theorem relies on the incompressibility of trajectories in phase space (Liouville's theorem) and microscopic time reversibility. 

\begin{figure}
\begin{center}
\leavevmode
\includegraphics[width=8.5cm,angle=0]{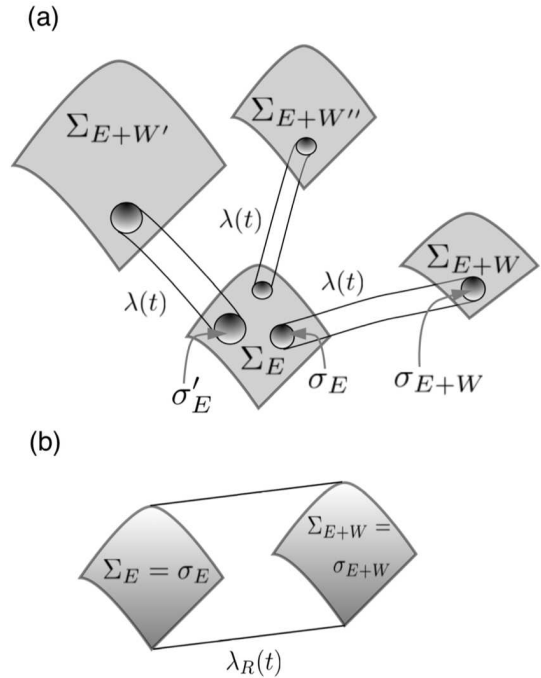}
\caption{Hamiltonian dynamics: A schematic illustration of the mapping of sets of states, initially with energy $E$, under the protocol $\lambda(t)$. (a) In an irreversible process, different amounts of work ({\it e.g.,} $W$, $W'$, and $W''$) may be performed, meaning that states from the manifold $\Sigma_{E}$ may be mapped to manifolds of different energies.   Note that due to the deterministic dynamics $\sigma_{E}$ and $\sigma'_{E}$ have no common elements and depend explicitly on the protocol $\lambda(t)$.  (b) For a reversible process, the manifold $\Sigma_{E}$ is mapped into the manifold $\Sigma_{{E+W}}$ where volumes of both the manifolds are same.}
\label{Ham}
\end{center}
\end{figure}

Consider the allowed phase-space manifolds $\Sigma_{E}$ and $\Sigma_{E+W}$ of the CS with energies $E$ and $E+W$ respectively (see Fig.\ref{Ham}). As the system is driven using the forward protocol $\lambda(t)$, a subset $\sigma_{E}$ of $\Sigma_{E}$ is mapped into a subset $\sigma_{E+W}$ of $\Sigma_{E+W}$ - for these trajectories an amount of work $W$ is performed. Other subsets of $\Sigma_{E}$ are mapped by the dynamics to other manifolds of different energies, for example $\Sigma_{E+W'}$, as indicated in the figure.  Denote the phase space volumes of $\sigma_{E}$ and $\sigma_{E+W}$ by $\omega_{E}$  and $\omega_{E+W}$ and those of  $\Sigma_{E}$ and $\Sigma_{E+W}$ by $\Omega_{E}$  and $\Omega_{E+W}$, respectively.  The incompressibility of phase space trajectories guaranties that $\omega_{E} = \omega_{E+W}$.  Microscopic time reversal symmetry (Condition 1) implies that if the driving protocol is reversed \cite{footnote-reverse}, then $\sigma^\dagger_{E+W}$ will be precisely mapped onto $\sigma^\dagger_{E}$, where $\sigma^\dagger_{E+W}$ and  $\sigma^\dagger_{E}$ are the time-reversed images (i.e., all momenta reversed) of $\sigma_{E+W}$ and $\sigma_{E}$, respectively. These clearly all have the same volume in phase space. In the special case when the protocol is a reversible process, the entire manifold $\Sigma_{E}$ will be mapped onto the entire manifold $\Sigma_{E+W}$, where $W$ is the energy added to the system in the reversible process, with no change in entropy.

Since in equilibrium all states on a constant energy manifold are equally probable, the probability of performing work $W$ in the forward protocol is given by $P_{F}(W) = \omega_{E}/\Omega_{E}$.  Similarly, for the reverse process $P_{R}(-W) = \omega_{E+W}/\Omega_{E+W}$.  We expect that in the limit of a large heat reservoir $E \gg W$, $P_{F}(W)$ and $P_{R}(-W)$ will not depend on $E$, which, together with $\omega_{E}=\omega_{E+W}$ gives
\begin{eqnarray}
P_{F}(W)/P_{R}(-W)&=&\Omega_{E+W}/\Omega_{E} \nonumber \\
&=& e^{S_{CS}(E+W)-S_{CS}(E)} \;.
\label{phase-space}
\end{eqnarray}
Here we have identified  $\Omega_{E+W} = e^{S_{CS}(E+W)}$ (and likewise for $\Omega_{E}$), where $S_{CS}(E+W)$ is the entropy of the CS with energy $E+W$.  We may now couch this result in the standard form (Eq. \ref{eq:CT}): First expand to first order in $W$ and use $\Delta F_{CS} = W - T \Delta S_{CS}$.  Next, use Condition 2 that  $F_{CS}=F_{TS}+F_{B}$, and Condition 3 that $F_{B}$ is independent of $\lambda$ (and so $\Delta F_{CS}=\Delta F_{TS}$).  

Note that for classical systems the condition $F_{CS}=F_{TS}+F_B$ is rather strong. It implies that either a thermodynamic limit of the test system is taken or that the interaction between the test system and the heat reservoir is weak. The latter would imply negligible local heating while the former would restrict the result to large test systems. If neither of these cases holds, we can not define the test system as a distinct object. In this case, as evident from the above proof, the Crooks' theorem still holds with $\Delta F_{TS}$ replaced by $\Delta F_{CS}$ implying that one has to consider the free-energy difference of the full combined system. Its value is expected to be dependent on the interaction energy between the test system and the heat reservoir. As we argue below, for stochastic systems these restrictions are less severe and one can have strong local heating even when $F_{CS}=F_{TS}+F_B$.

\section{Stochastic Dynamics}

\begin{figure}
\begin{center}
\leavevmode
\includegraphics[width=8.5cm,angle=0]{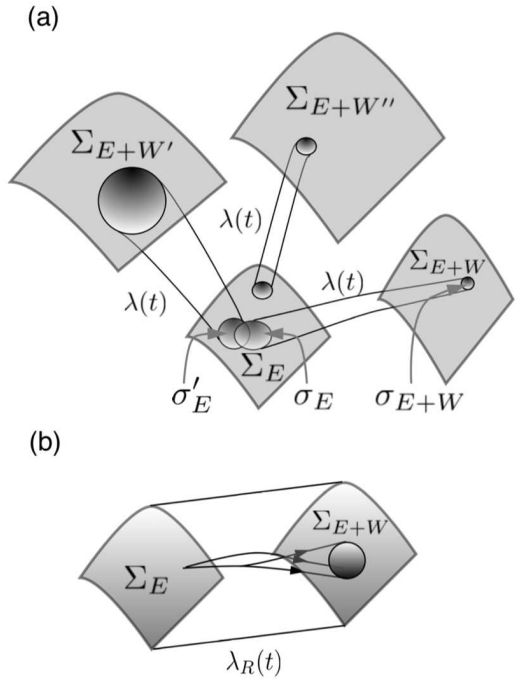}
\caption{Stochastic dynamics: A schematic illustration of the mapping of sets of states, initially with energy $E$, under the protocol $\lambda(t)$.  In this case, the dynamics do not map a given initial to a single final state.  Time-reversibility only implies that a forward trajectory has a corresponding time-reversed trajectory of equal statistical weight.  (a) In an irreversible process, different amounts of work ({\it e.g.,} $W$, $W'$, and $W''$) may be performed, meaning that states from the manifold $\Sigma_{E}$ may be mapped to manifolds of different energies.   In this case, $\sigma_{E}$ and $\sigma'_{E}$ may have common elements.  (b) For a reversible process, the manifold $\Sigma_{E}$ is mapped into the manifold $\Sigma_{{E+W}}$.  This notwithstanding, a given initial state may evolve along many different trajectories $Y(t)$, as indicated.}
\label{Stoch}
\end{center}
\end{figure}

Stochastic dynamics arise either due to incomplete knowledge of some degrees of freedom of the system and the heat reservoir or due to quantum effects (or both). Although the choice of an effective stochastic dynamics may be justified from the underlying microscopic classical or quantum dynamics \cite{vanKampen, Tolman, Haar}, it is not obvious that the fluctuation theorems are valid when only the effective stochastic dynamics is provided. For example, noise of quantum origin (or in a more general purely mathematical settings) is not captured by Hamiltonian dynamics and thus invalidates the proof for the Hamiltonian case.
The central difference between the stochastic case and that of classical Hamiltonian dynamics is that here the evolution of a volume in phase space is not incompressible.  This means that a given initial state may evolve under the dynamics into many final states. Employing the same notation as in section II, this difference is illustrated in Figure \ref{Stoch}, which shows that the volume $\omega_{E}$ is not, in general, equal to $\omega_{E+W}$. 

\subsection{Crooks' Theorem}

We now turn to the proof of the Crooks' theorem for a stochastic time reversible dynamics. To understand irreversible and dissipative physics in a stochastic settings, there are various ways one can model a system \cite{Blythe}. We do in the following fashion. We consider an isolated CS comprised of a test system coupled to a reservoir.  Per Condition 1, we assume that the CS evolves through Markovian dynamics obeying microscopic time reversibility.  Formally, this means that the transition rate $w_{j,i}$ ($w_{i,j}$) from configuration $i$ to $j$ ($j$ to $i$) of the CS satisfies
\begin{equation}
w_{j,i} = w_{i,j} \; ;
\label{DB}
\end{equation}
This is just the statement of detailed balance for an isolated system \cite{vanKampen}.
We stress that this does not imply Eq. \ref{CanonicalDB}, that is, the {\it effective} heat reservoir is neither Markovian nor in quasi-equilibrium, and the above condition is a much weaker assumption than that made in References \cite{CrooksPRE1999, CrooksJStatPhys1998}. 
   
Let us now consider the `forward' process in the time range $(-\infty < t < \infty)$,  where $\lambda(t) = \lambda_{A}$ for $(-\infty,0)$, $\lambda_{B}$ for $(\tau,\infty)$, and varies in a fixed fashion in $(0,\tau)$. At $t=-\infty$ the CS is in  an equilibrium state ${\cal A}$, and at $t=\infty$ the CS comes to an equilibrium state $\cal B$.  For a given protocol $\lambda(t)$ we indicate the probability of a given {\it trajectory} $Y(t)$ of the CS in the forward process by the functional ${\cal P}_F[Y(t),\lambda(t)]$. $Y(t) = \{x(t),X(t)\}$ includes all the degrees of freedom of the CS, that is those of the test system ($x(t)$) and of the heat reservoir ($X(t)$). The functional  ${\cal P}_R [\tilde Y(t),\tilde \lambda(t)]$ denotes the probability of the corresponding time-reversed trajectory $\tilde Y(t)=Y(-t)$ and time reversed forcing protocol,  $\tilde \lambda(t)=\lambda(-t)$, in which the CS begins in state $\cal B$ at $t=-\infty$ and ends in $\cal A$ at $t=\infty$. Eq. \ref{DB} implies that
\begin{equation}
{\cal P}_F [Y(t),\lambda(t)]= {\cal P}_R [\tilde Y(t),\tilde \lambda(t)]
\label{MR}
\end{equation}
for any trajectory. This simply says that the probability of any trajectory in the configuration space of the combined system is equal to the probability of its' corresponding time reversed trajectory. Work performed is defined as the energy difference between the two ends of the trajectory, i.e., $[H_{CS}(Y_{f}, \lambda_f) - H_{CS}(Y_{i}, \lambda_i)]$ where $H_{CS}(Y(t), \lambda(t))$ is the total energy of the combined system at any time $t$, subscript $i$ and $f$ denote the initial and final points of the trajectory. We denote the functional ${\cal W}_{F}[Y(t),\lambda(t)]$ as the work performed along the forward trajectory, and ${\cal W_{R}}[\tilde Y(t),\tilde \lambda(t)]$ is that for the corresponding time-reversed trajectory, with ${\cal W}_{F}[Y(t),\lambda(t)]=-{\cal W}_{R}[\tilde Y(t),\tilde \lambda(t)]$. The probability that an amount of work $W$ is performed in the forward and reverse protocols, respectively, are given by 
\begin{equation}
 P_F(W) = \\ 
\int P^{eq}_{\cal A}({Y_{-\infty}}) {\cal P}_F[Y,\lambda] \delta({\cal W}_F-W){\cal D}Y
\label{Crooks1}
\end{equation}
and
\begin{equation}
 P_R(W) = \\ 
\int P^{eq}_{\cal B}({\tilde Y_{\infty}}) {\cal P}_R [\tilde Y,\tilde \lambda] \delta({\cal W}_R-W){\cal D}{\tilde Y}
\label{CrooksR}
\end{equation}
where ${\cal D}Y \equiv \prod_{n=-\infty}^{\infty} dY_n$ denotes the integration over all trajectories in the configuration space,
$P^{eq}_{\cal A}({Y_{-\infty}})$ and $P^{eq}_{\cal B}({Y_{\infty}})$ are the initial and final equilibrium distributions of the CS, and we have omitted the explicit trajectory dependence of ${\cal W}_{F}[Y,\lambda]$ and ${\cal W}_{R}[\tilde Y,\tilde \lambda]$.

The CS is in equilibrium for $t=\pm\infty$, and an amount of work $W$ is performed only during the interval $(0,\tau)$.  For $t=\pm\infty$ the CS is described by a microcanonical ensemble, and the probability of any microstate $Y = \{x,X\}$ is the inverse of the total number of microstates, $e^{-S_{CS}(t=\pm\infty)}$. Now we expand $S_{CS}(E+W)$ to first order in $W$ and use $\Delta F_{CS} = W - T \Delta S_{CS}$ where the inverse temperature $T^{-1}=\partial S_{CS}(E)/\partial E$.  According to Condition 3, {$\lambda$} couples directly only to the test system, while Condition 2 implies that the free energy of the CS is the sum of the free energies of the reservoir and test system.  Thus we have that 
\be
\frac{P_{\cal A}^{eq}(Y_{-\infty})}{P_{\cal B}^{eq}(Y_{\infty}) } = e^{\beta ({\cal W}_F - \Delta F_{TS})},
\label{ratio_prob}
\ee 
where ${\cal W}_F$ is the work performed along the forward path and $\Delta F_{TS} = F_{TS}({\cal B})-F_{TS}({\cal A})$ is the free-energy difference of the test system at $t=\infty$ and $t=-\infty$, that is, for the values $\lambda_{B}$ and $\lambda_{A}$ \cite{HG}. Putting Eq. \ref{MR} and Eq. \ref{ratio_prob} in Eq. \ref{Crooks1}, and using ${\cal W}_F=-{\cal W}_R$, we get
\bea
\nonumber P_F(W) = \int e^{-\beta ({\cal W}_R+\Delta F_{TS})} P^{eq}_{\cal B}({Y_{\infty}}) {\cal P}_R[Y,\lambda] \mbox{~~~~~~~~~~~~~} 
\\ 
\nonumber \times \delta({\cal W}_R+W){\cal D}Y
\\
\nonumber \mbox{~~~} = e^{\beta (W-\Delta F_{TS})} \int P^{eq}_{\cal B}({Y_{\infty}}) {\cal P}_R[\tilde{Y},\tilde{\lambda}] \delta({\cal W}_R+W){\cal D}Y
\\
=e^{\beta (W-\Delta F_{TS})} P_R(-W) \mbox{~~~~~~~~~~~~~~~~~~~~~~~~~~~~~~~~~~~}
\eea
which is obtained after comparing with Eq. \ref{CrooksR}. This proves the Crooks' theorem for stochastic dynamics. Note that also here we assume that $P_F(W)$ and $P_R(W)$ are independent of $E$ in the limit of large heat reservoir. Finally, where Conditions 2 and 3 do not apply similar considerations to those discussed for Hamiltonian dynamics hold. 

Note that the conditions on the validity of Crooks' theorem are weaker here than for Hamiltonian dynamics. For Hamiltonian dynamics, Liouville's theorem and Condition 1 imply time reversibility in the sense that each microstate in the initial manifold maps under the action of $\lambda(t)$ onto a single final microstate, and under the reverse protocol the final microstate, with all momenta reversed, in turn maps back to the original microstate with all momenta reversed. In the stochastic case the constraint is weaker in the sense that an initial microstate may end up in any one of many final  microstates; all that is required for the proof is that the weights of path connecting states in the forward and reverse direction are equal. 

In contrast to classical Hamiltonian dynamics, for stochastic dynamics there is no connection between local heating and the interactions between the test-system and the heat reservoir. Indeed, the energy function of the stochastic system may contain no interaction term between the test system and the heat reservoir. However, this does not mean that there is no heat transfer, and consequently local heating can occur. This is illustrated in the numerical example of Sec. III.

\subsection{Jarzynski Equality}

Although the Jarzynski equality can be derived by direct integration of Crooks' Theorem, it is instructive to present a proof for stochastic dynamics in the spirit of the previous section.  We discretize the process and consider a general protocol $\{ \lambda_0, \lambda_1, \dots,\lambda_{n_{\lambda}}\}$, where $\lambda_k$ is the value of $\lambda$ for the $k^{th}$ update interval. The process consists of changing the value of $\lambda$ (while leaving $Y = \{x,X\}$ unchanged), updating the CS, changing $\lambda$ again, etc.  Denoting the values of the degrees of freedom of the CS at step $k$ by $Y_{k}\equiv (x_{k}, X_{k})$, this process is specified by the sequence of transitions  
$(Y_0, \lambda_0) \rightarrow (Y_0, \lambda_1) \rightarrow  (Y_1, \lambda_1) \rightarrow (Y_1, \lambda_2) \dots$ .
The transition probability matrix between the states ($Y_{k-1},\lambda_{k}$) and ($Y_{k},\lambda_{k}$), which has an implicit dependence on time through the changing parameter $\lambda$, is denoted by ${\overline w}_{k,k-1}$.  Condition 1, with the Markovian dynamics for the entire CS implies that 
\begin{equation}
\int {\overline w}_{k+1,k} P^{eq}_{CS}(Y_k,\lambda)dY_k =P^{eq}_{CS}(Y_{k+1},\lambda)\;.
\label{eq:JElemma}
\end{equation}
(Condition 1 guarantees that dynamics for a given value of $\lambda$ only connects microstates in the equilibrium ensemble.)

Conditions 2 and 3 imply that the work at time step $k$ is defined as $\Delta W_k = H(x_k, \lambda_{k+1}) - H(x_k, \lambda_{k})$ where $H(x, \lambda)$ is the energy of the test system in microstate $x$ at parameter value $\lambda$.  The total work performed in the process is thus given by $W = \sum_{k=1}^{n_{\lambda}} \Delta W_k$.  Condition 2 implies that
\begin{equation}
	P^{eq}_{TS}(x,\lambda)=\exp(-\beta H(x, \lambda))/{\cal Z} \;,
	\label{eq:sysP}
\end{equation}
where ${\cal Z}$ is the partition function of the test system.
From Eqs.  \ref{eq:JElemma} and \ref{eq:sysP}, we have
\begin{equation}
\nonumber \int e^{- \beta \Delta W_k} {\overline w}_{k+1,k} P^{eq}_{CS}(Y_k)  dY_k =
 \frac{{\cal Z}_{k+1}}{{\cal Z}_k} P^{eq}_{CS}(Y_{k+1}) \; ,
\label{relation1}
\end{equation}
where ${\cal Z}_k$ is defined in Eq. \ref{eq:sysP} with $\lambda=\lambda_k$.
Using this repeatedly in the relation 
\begin{eqnarray}
\nonumber \langle e^{-\beta W} \rangle = \int P^{eq}_{CS}(Y_0,\lambda_{0}) \prod_{k=0}^{n_{\lambda}} e^{- \beta \Delta W_k} {\overline w}_{k+1,k}  dY_k \; ,
\label{Avg_exp}
\end{eqnarray}
we readily obtain Eq. \ref{eq:JE} by identifying $F_k=-\beta^{-1} \ln ({\cal Z}_k)$, the free energy of the test system for $\lambda=\lambda_k$.

\section{Numerical Model}

In the preceding sections we have shown that local heating does not invalidate either the Crooks' Theorem or the Jarzynski Equality.  This notwithstanding, local heating has a crucial effect on the convergence of the results.  We illustrate these results using a simple stochastic model.  We first show numerically the validity of both the Jarzynski equality and the Crooks' theorem for stochastic dynamics respecting Eq. \ref{DB}. This allows us to study the effect of local heating on the typical number of experimental runs needed to estimate the average in Eq. \ref{eq:JE}.  It is shown that the larger the local heating, the greater the number of runs needed. 

\begin{figure}
\begin{center}
\leavevmode
\includegraphics[width=8.5cm,angle=0]{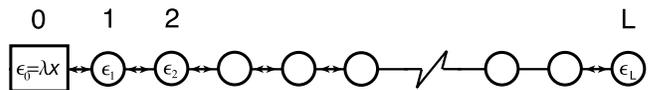}
\caption{Schematic diagram of the CS. The test system occupies site $i=0$, and the sites of the heat reservoir are labeled by $i=1,2,3 \dots L $ where $L$ is very large. Interactions are between neighboring sites.}
\label{SB}
\end{center}
\end{figure}

Consider a one-dimensional \cite{footnote3} lattice of $L+1$ sites, with sites labeled $i=0,...,L$; this is our CS. The test system is placed at site $i=0$, and the heat reservoir occupies sites $1 \leq i \leq L$, as indicated in Fig. \ref{SB}. At each site we define a real and positive energy variable $\epsilon_i$. For the test system we assume, for simplicity, a single degree of freedom denoted by $x$ which takes values $x \geq 0$. The energy of the test system is chosen to have the simple form $\epsilon_0=H(x,\lambda)=\lambda x$, where $\lambda$ is the external forcing parameter.  (Similar results were obtained for different energy functions, such as $\epsilon_0=H(x,\lambda)=\lambda x^{2}$.)

The parameter $\lambda$ is updated according to a set deterministic protocol $\lambda(t)$ in the time interval $(0, \tau)$.  During this time, the CS evolves with random sequential stochastic dynamics, with each site $0 \leq i < L$ in the CS interacting with an arbitrary rate $r$ with the site to its right by stochastically redistributing their total energy:  
\begin{equation}
\epsilon^{\prime}_i = q (\epsilon_i + \epsilon_{i+1})  \mbox{~~~~~and~~~~~}  \epsilon^{\prime}_{i+1} = (1-q) (\epsilon_i + \epsilon_{i+1}) \;.
\label{rule1}
\end{equation}
Here $0 \leq q \leq 1$ is drawn from a uniform distribution and the prime denotes the values after redistribution \cite{KPM}. The update rule for $i=0$ (the test system) and the role of parameter $\lambda$ in the dynamics are explained in more detail as follows. Since the energy of the test system $\epsilon_0= \lambda x$, the maximum possible value of $x$ in the next update step is $x_{max}=(\epsilon_0+\epsilon_1)/\lambda$. Now we generate a random number $x_r$ which lies uniformly between $0$ to $x_{max}$ and the update rules for the sites $0$ and $1$ are: $\epsilon_0^{\prime} = \lambda x_r$ and $\epsilon^{\prime}_1 = (\epsilon_0+\epsilon_1-\lambda x_r)$ where $\epsilon^{\prime}_0$ and $\epsilon^{\prime}_1$ are the energies of the respective sites after redistribution. In this way we ensure that the transition to any of the states, arising due to the energy exchange, are equally likely. One can easily check that the updating of sites $i=0$ and $i=1$ agrees with Eq. \ref{rule1} for this particular model.  
In accordance with standard definitions \cite{JarzynskiPRE1997} the work done on the test system is given by
\begin{equation}
	W=\int_0^\tau \frac{\partial H}{\partial \lambda} \frac{d \lambda}{d t} dt = \int_0^\tau x(t) \frac{d \lambda}{d t} dt \;.
	\label{eq:work}
\end{equation}
Large $r$ compared to $d \lambda/ dt$, the rate of change of external parameter, means a quickly thermalizing reservoir; in the limit $r \rightarrow \infty$ for any finite $d \lambda/ dt$, we expect the reservoir to lose all memory and become Markovian.

When $\lambda(t)=\lambda_0 = constant$, we update the test system and the reservoir by repeatedly using the dynamics in Eq. \ref{rule1}. Clearly the total energy of the CS is conserved, and the dynamics satisfy detailed balance with respect to a measure that is uniform on a constant energy surface, as required in a micro-canonical ensemble.  It is straightforward to show that in the thermodynamic limit $L\to \infty$, the equilibrium energy distribution of any site $i=0, \ldots L$ takes the form
 \begin{equation}
P(\epsilon_{i}) =  \beta e^{-\beta \epsilon_i}
\label{ansatz}
\end{equation}
where $\beta^{-1}=\sum_{i=0}^{L} \epsilon_i/(L+1)$.  The distribution of the CS is a product measure: $P_{CS}(\{\epsilon_j\}) = \prod_{i} \beta e^{-\beta \epsilon_i}$.  This allows us to calculate the partition function of the test system for a fixed value of $\lambda$: ${\cal Z} (\lambda) = \int_0^{\infty} e^{- \beta \lambda x} dx = (\beta \lambda)^{-1}$ with the free-energy given by $F(\lambda) = -\beta^{-1} \ln {\cal Z}$.

%\section{3. Simulation results}

As the test system is driven, there is local heating in the region of its contact with the reservoir \cite{footnote2}, with the consequence that the energy distributions of the reservoir sites near to the test system (e.g. $i=1, 2, 3$ etc) deviate significantly from the equilibrium distribution as given in Eq. \ref{ansatz}.  We study this numerically by employing the following driving protocol $\lambda(t)$: Starting at $\lambda(0)=\lambda_0$ we increase the forcing parameter to $\lambda(\tau)\equiv\lambda_0+\Delta \lambda$ at intervals $\delta \tau$ in $n_{\lambda}\equiv\tau/\delta \tau$ discrete steps of size $\delta \lambda \equiv \Delta \lambda /n_{\lambda}$. Between these updates the dynamics specified in Eq. \ref{rule1} are carried out using standard Monte-Carlo methods. By discretizing Eq. \ref{eq:work}, the work performed due to changing the parameter from $\lambda_k$ to $\lambda_k + \delta \lambda$ at $k$-th update step is defined as $\Delta W_k = H(x_k, \lambda_k + \delta \lambda) - H(x_k, \lambda_k)$ where $x_k$ is the test system degree of freedom. The total work performed is $W=\sum_{k=1}^{n_\lambda} \Delta W_k$.

For fixed $\Delta \lambda$, a {\it reversible} process occurs in the limit $\frac{\lambda(t+\delta t) - \lambda(t)}{r \delta t} \to 0$.  For this to occur, the limit $n_{\lambda} \rightarrow \infty$ must be taken first, followed by either taking $r \to \infty$ or $\tau \to \infty$. {\it All other protocols, in particular those with finite $n_{\lambda}$, are irreversible}. In the following we compare systems by putting $r=1$ (which sets the time scale) and varying the time interval $\tau$.

\begin{figure}
\begin{center}
\leavevmode
\includegraphics[width=8.4cm,angle=0]{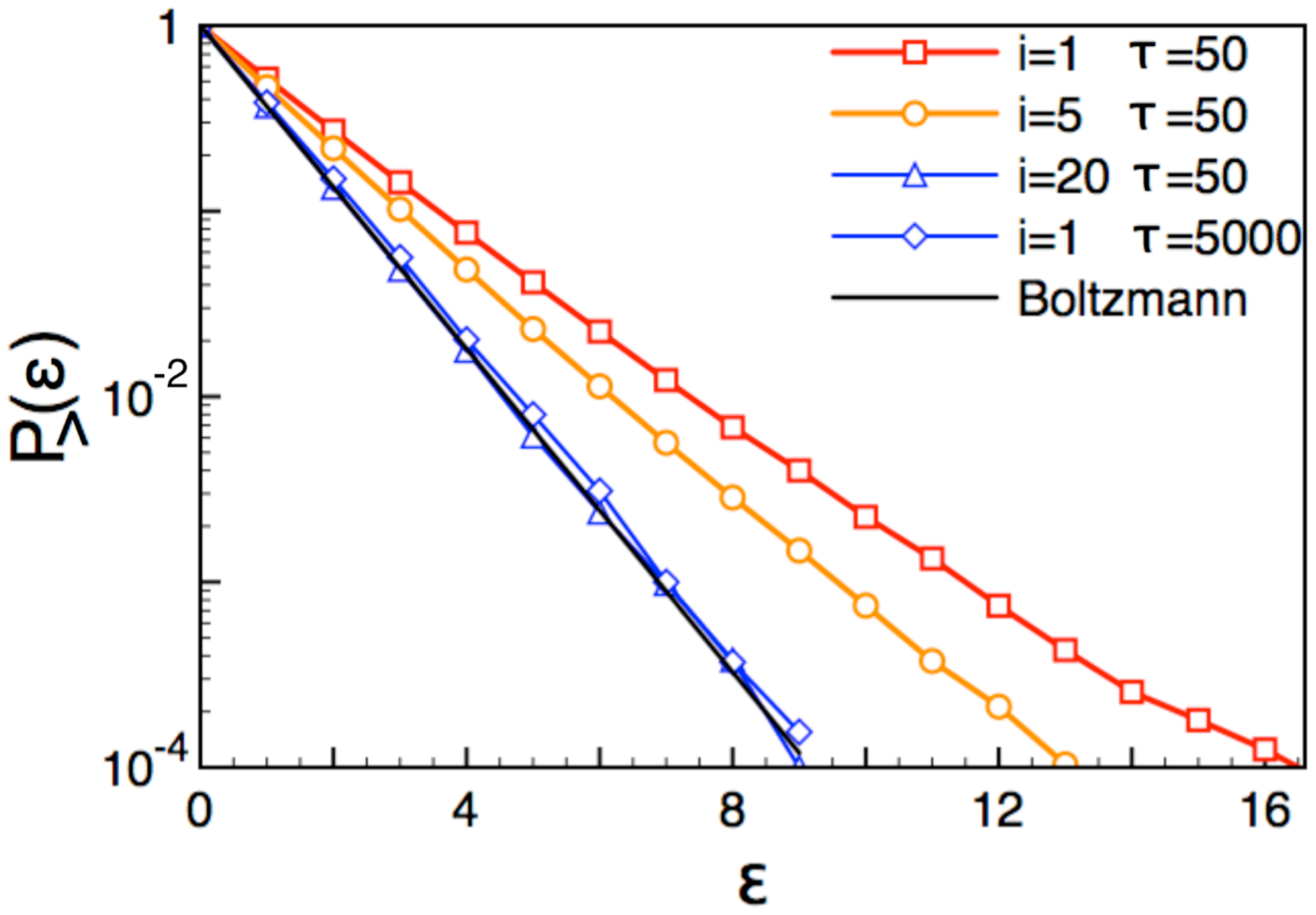}
\includegraphics[width=7.4cm,angle=0]{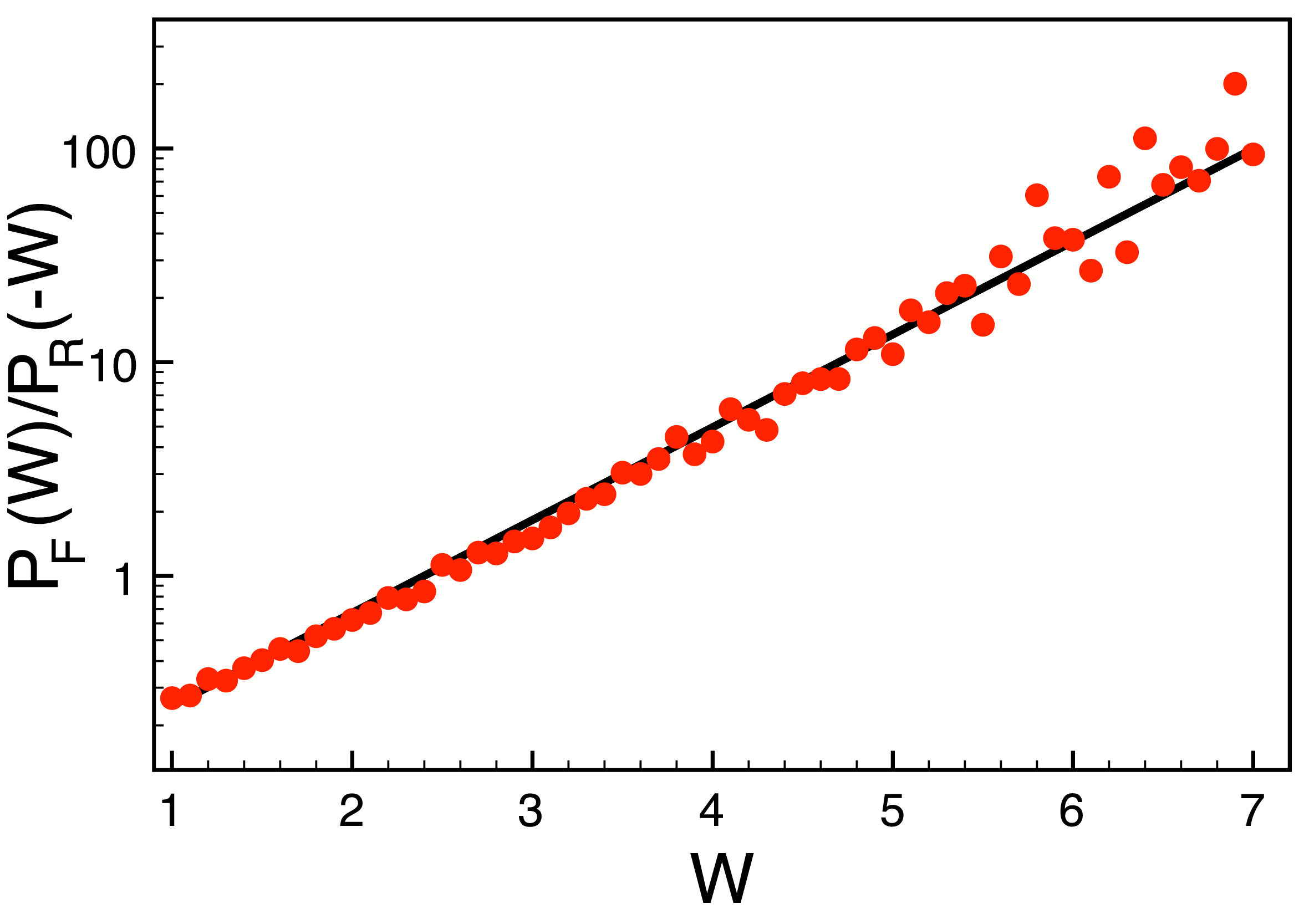}
\caption{Top panel: The integrated final $(t=\tau)$ energy distributions, $P_>(\epsilon)$ for several sites, $i$ for a short ($\tau=50$) and long ($\tau=5000$) experimental time. Bottom panel: The solid line is $\exp(\beta W-\beta \Delta F)$, where $\Delta F = \beta^{-1} \ln [(\lambda_0+\Delta \lambda)/\lambda]$, obtained analytically. We take $r=1$ in an arbitrary unit and measure the time with respect to it. Results are obtained for $L=100$, $T=1$, $\lambda_0=1$, $\delta \lambda=1$ and $n_\lambda=10$, and by averaging over $10^{5}$ realizations.}
\label{non-Boltzmann}
\end{center}
\end{figure}

The effect of local heating on the single-site energy distribution is shown in Fig. \ref{non-Boltzmann}, where we plot the (integrated) distributions $P_{>}(\epsilon) = \int_{\epsilon}^{\infty}P(\epsilon')d\epsilon'$ for several sites for two values of $\tau$ and with the same protocol for $\lambda(t)$. We see that far from the test system ($i \gg 1$) the reservoir is Boltzmann-distributed (as in Eq. \ref{ansatz}), but there is significant deviation near to the point of contact with the test system.  As expected, for large $\tau$ this effect decreases.  In a sense, local heating is a measure of the non-Markovian nature of the reservoir as experienced by the test system. This notwithstanding, Fig. \ref{non-Boltzmann} demonstrates that even in the presence of local heating, the Crooks' theorem is clearly satisfied. The validity of the Crooks' theorem automatically implies that the Jarzynski equality also holds.   

\begin{figure}
\begin{center}
\leavevmode
\includegraphics[width=7.5cm,angle=0]{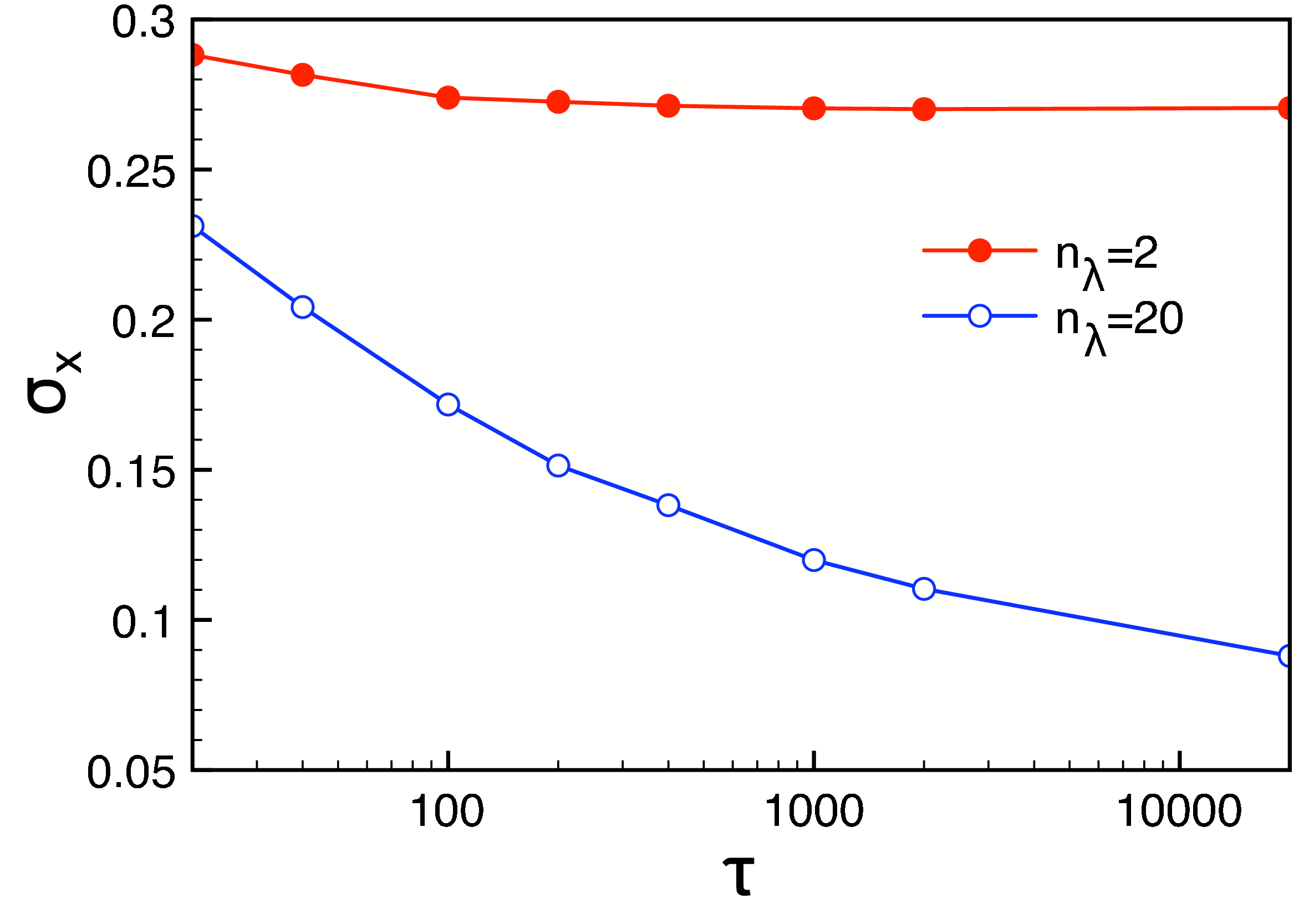}
\includegraphics[width=7.5cm,angle=0]{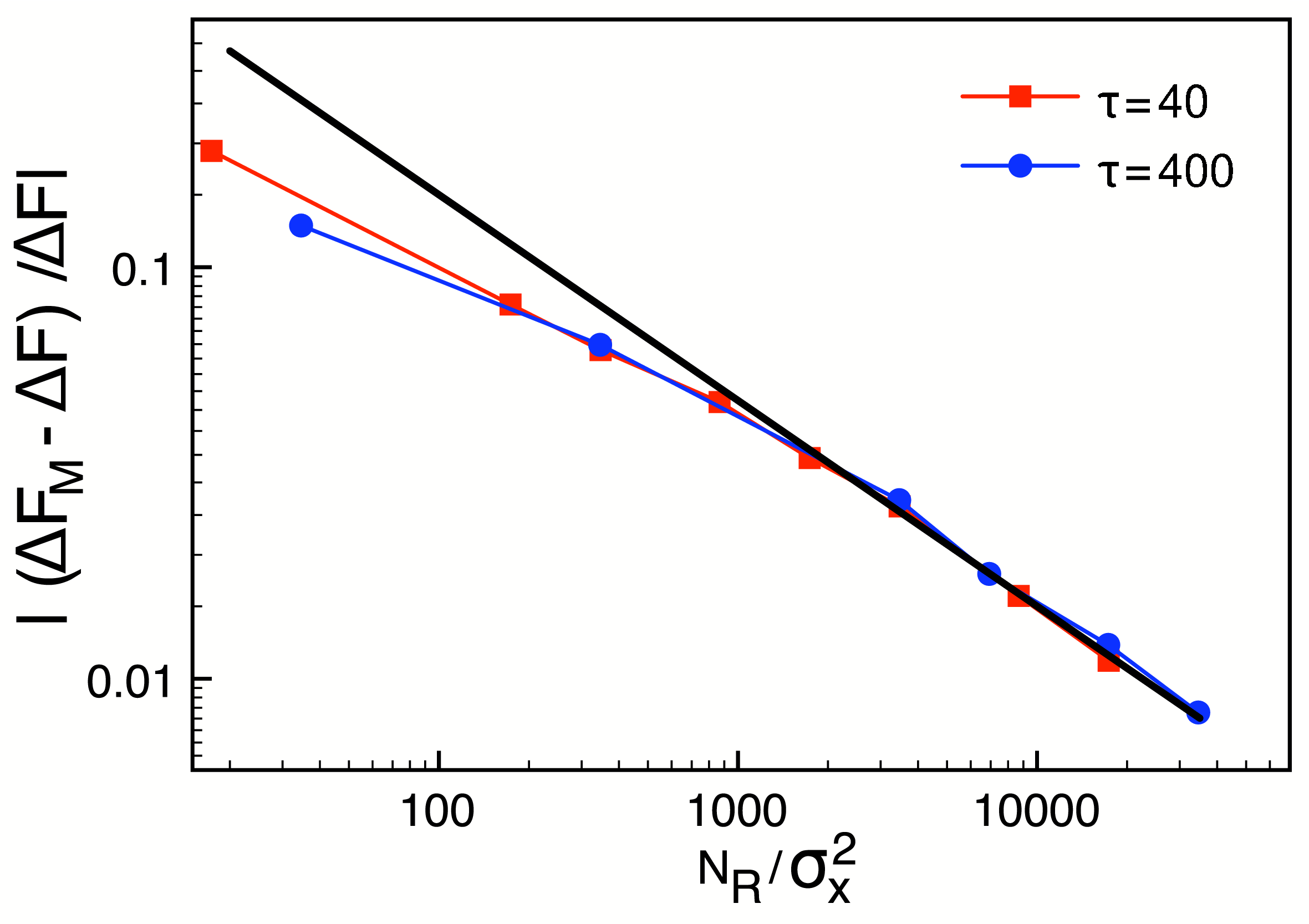}
\caption{Top panel: $\sigma_{X}$ as a function of $\tau$, and $N_{R}=10^4$. Bottom panel: The fractional error in the measured free energy difference $\Delta F_M$, versus $N_{R}/\sigma_{X}^{2}$ where averaging is done over $10^3$ realizations for each value of $N_R/\sigma_X^2$, and the straight line has slope $-0.5$. Results are for $L=100$, $T=4$, $\lambda_0=2$, $\Delta \lambda=1$.}
\label{sigmaX}
\end{center}
\end{figure}

One of the chief uses of the Crooks' theorem and Jarzynski equality is to experimentally measure the (relative) free energy of a test system.  For this purpose, local heating is of central importance, since it has a pronounced effect on the number of experimental runs which must be carried out to accurately determine the free-energy difference $\Delta F = F_{\cal B}-F_{\cal A}$ \cite{Presse}.  Our discussion will focus on the Jarzynski equality, but similar conclusions hold for the Crooks' theorem.  In an experiment, the quantity $e^{-\beta W}$ is measured once per run.  The experimental average over $N_{R}$ runs is thus 
$$
\overline{e^{-\beta W}} = \frac{1}{N_R}\sum_{l=1}^{N_{R}}e^{-\beta W_{l}} \; .
$$
where $W_l$ is the work performed in the $l$-th experimental run. Assuming the validity of the Jarzynski equality, this gives an estimate of the free energy difference measured from the experiment:
$$
\beta \Delta F_M = -\ln {\overline{e^{-\beta W}}}  \; .
$$
As we show, the more pronounced the local heating, the larger $N_{R}$ must be in order for $\Delta F_M$ to accurately estimate the actual free energy difference. To see this, note that the scale of the estimated error of the average of $X\equiv e^{-\beta (W-\Delta F)}$
is set by the width $\sigma_X \equiv \sqrt{\langle X^2\rangle - \langle X \rangle ^2}$ of the probability distribution $P(X)$ \cite{Gore}. 
The central limit theorem states that for $N_{R}$ runs, the error in the estimate of $\langle X \rangle$ scales as $\sigma_X N_{R}^{-1/2}$ for large $N_{R}$. It is straightforward to show that the error in the estimate of  $\Delta F$ scales in the same fashion. Therefore, the smaller $\sigma_X$, the fewer the runs needed to obtain an accurate estimate of $\Delta F$.  This is especially relevant when the experimental protocol drives the system in a highly irreversible fashion.

Fig. \ref{sigmaX} shows the dependence of $\sigma_{X}$ on $\tau$ for two values of $n_\lambda$. As is clearly seen in the data, the longer the time of the experiment, the smaller $\sigma_X$.  In the absence of local heating, such as occurs in the $\tau \to  \infty$ limit, $\sigma_{X}$ achieves a parameter-dependent constant value, as seen in Fig. \ref{sigmaX}.  We note that since $n_\lambda$ is finite, the process is irreversible even for infinitely long experiments, with the consequence that $\sigma_X$ does not reach zero.  However, $\sigma_X$ diminishes with increasing $n_\lambda$, tending to zero in the reversible limit discussed above.  Fig. \ref{sigmaX} shows the deviation, $\Delta F_M - \Delta F$, of the estimated free-energy difference from the actual value as a function of the expected scaling variable $N_R/\sigma_X^2$. Once again, we see that the larger the local heating, the more experimental runs are needed to evaluate the free-energy. Note that by considering a system with maximal local-heating, an upper-bound on the number of realizations can be obtained.  This can be realized by making sudden changes in the test system, thus effectively disconnecting the heat reservoir after an initial equilibrium distribution has been attained.

\begin{acknowledgments}
We are grateful to Richard Blythe for many valuable suggestions and comments. YK and DL acknowledge support from grants 1183/06 and 660/05 of the Israel Science Foundation. PP acknowledges a fellowship of the Israel Council for Higher Education. 
\end{acknowledgments}


\begin{thebibliography}{99}
\bibitem{footnote-reverse} In the presence of a field which breaks time-reversal symmetry (such as a magnetic field) equivalent theorems may be derived where one reverses the direction of the field during the backward protocol.
\bibitem{HG} Note that Eq. \ref{ratio_prob} is valid for the free-energy consistent with the given constraints, e.g., the free-energy $F_{TS}$ may be either Helmohltz' free-energy or Gibbs' free-energy depending on the constant temperature-volume processes or constant temperature-pressure processes, respectively.
\bibitem{footnote3} The effect of local heating is expected to be largest in 1D, since heat diffuses less effectively than in higher dimensions.
\bibitem{footnote2} We take the CS large enough so that the local heating has a negligible effect on the total energy of the CS.
\end{thebibliography}
\end{document}